\documentclass[preprintnumbers,amsmath,11pt,amssymb,floatfix,superscriptaddress,nofootinbib]{article}

\def\mytitle#1{\setcounter{equation}{0}
	\setcounter{footnote}{0}
	\begin{flushleft}\Large\textbf{#1}\end{flushleft}
	\vspace{0.25cm}}
\def\myname#1{\leftline{{\large #1}}\vspace{-0.13cm}}
\def\myplace#1#2{\small\begin{flushleft}\textit{#1}\\
		\texttt{#2}\end{flushleft}}

\usepackage{graphicx}% Include figure files
\usepackage{dcolumn}% Align table columns on decimal point
\usepackage{color}
\usepackage{bm}% bold math
\usepackage{amsmath}
\usepackage{amsfonts}
\usepackage{amssymb}

\input epsf

\begin{document}

\mytitle{A Study of Universal Thermodynamics in Lanczos-Lovelock gravity}
\myname{Saugata Mitra\footnote{saugatamitra20@gmail.com}}
\myplace{Department of Mathematics, Jadavpur University, Kolkata 700032, West Bengal, India.}
\author{Subhajit Saha\footnote {subhajit1729@gmail.com}}
\myplace{Department of Mathematics, Jadavpur University, Kolkata 700032, West Bengal, India.}
\author{Subenoy Chakraborty\footnote {schakraborty.math@gmail.com}}
\myplace{Department of Mathematics, Jadavpur University, Kolkata 700032, West Bengal, India.}{}
%%%%%%%%%%%%%%%%%%%%%%%%%%%%%%%%%%%%%%%%%%%%%%%%%%%%%%%%%%%%%%%%%%%%%%%%%%%%%%%%%%%%%%%%%%%%%%%%%%%%%%%%%%%%%%%%%%%%%%%%%%%%
\begin{abstract}

A study of Universal thermodynamics is done in Lanczos-Lovelock gravity. The Universe is chosen as FRW model bounded by apparent or event horizon. The unified first law is examined, assuming extended Hawking temperature on the horizon. As a result there is a modification of Bekenstein entropy on the horizons. Further the validity of generalized second law of thermodynamics and thermodynamical equilibrium are also investigated for perfect fluid (with constant equation of state), modified chaplygin gas model and holographic dark energy model.\\

\end{abstract}
\section{Introduction}
The field equations in the four dimensional Einstein gravity can be obtained from the Einstein-Hilbert action which is linear in the Riemann tensor. These equations of motion determine how the matter fields act as a source for curvature. Unfortunately, there is no elegant principle to determine these field equations of gravity. It is therefore reasonable to study the most general system of field equations which are second order in the independent variables and which satisfy the criteria of general covariance. Further, it can also be speculated that the Einstein-Hilbert action, which is only an effective gravitational action valid for small curvatures at low energies, will certainly undergo a modification by higher derivative interactions in any attempt to perturbatively quantize gravity as a field theory \cite{a1.1,a1.2,a1.3,a1.4,a1.5,a1.6,a1.7,a1.8,a1.9}. These observations lead us to a class of models first proposed by Lanczos \cite{a2} and Lovelock \cite{r16}. The field equations for these models can be derived from an action functional in D-dimensions \cite{a2,r16,a4} which is a polynomial in the curvature tensor. When D=4, the action becomes the standard Einstein-Hilbert action (for a recent review on Lanczos-Lovelock gravity, see \cite{a5} and references therein). The quasi linearity feature of the Einstein-Lovelock equations of motion guarantees that they can be formulated as a Cauchy problem with some constraints on the initial data \cite{a6} and also renders the quantization of the linearized Lanczos-Lovelock theory as ghost-free \cite{a7,a8,a9}.

Recent investigations have revealed that the Lanczos-Lovelock models are motivated by the intimate connection between gravitational dynamics and horizon thermodynamics. This connection was established by the pioneering work of Bekenstein, Hawking, Davies, Unruh and others showing that horizons in general possess thermodynamic properties like entropy \cite{a10,a11} and temperature \cite{a12,a13,a14,a15}. The gravitational field equations in Einstein and in more general gravity theories \cite{a16,a17,a18} as well as in all the Lanczos-Lovelock models can be derived from a thermodynamical extremum principle \cite{a5} and the action functional itself can be given a thermodynamic interpretation. Consequently, the emergent gravity paradigm (for a review, see \cite{a16,a17,a18,a19} and references therein) which enables us to think of space-time as some sort of fluid with its thermodynamic properties arising from the dynamics of underlying "atoms of space-time", has gained momentum in recent years. Such studies have also unravelled several peculiar geometrical features of Lanczos-Lovelock models \cite{a5,a20}.

In the present work, universal thermodynamics in Lanczos-Lovelock gravity has been studied using extended Hawking temperature on the horizon (event/ apparent) and modified horizon entropy. The validity of the generalized second law of thermodynamics (GSLT) and thermodynamical equilibrium (TE) have been examined for three choices of the cosmic fluid namely,\\
i) perfect fluid with constant equation of state,\\
ii) the modified Chaplygin gas,\\
and iii) the holographic dark energy.\\
In each case, the first and the second derivatives of the entropy on the horizon (event/ apparent) have been expressed and due to complicated expressions, they have been plotted against relevant parameters in the respective fluid models, for different space-time dimensions. In the holographic dark energy case, the values for the parameters c and $\Omega_d$ have been obtained from three Planck data sets presented in Table 2. Finally, the constraints on the parameters involved have been presented in tabular form.

\section{A brief overview of Lanczos-Lovelock gravity}

The Lagrangian density of Lanczos-Lovelock gravity consists of the dimensionally extended Euler densities which are given by \cite{r16,b1,b2,r10,r17,r18}
\begin{equation}
\label{action}
L=\sum_{i=0}^{[\frac{D}{2}]}c_iL_i
\end{equation}
where $c_i$'s are constants, D is the space-time dimension  and $L_i$'s are the Euler densities of 2i-dimensional manifolds and are defined by
\begin{equation}\label{Z1}
L_i=\frac{1}{2^i}\delta^{a_1b_1 \ldots a_ib_i}_{c_1d_1 \ldots c_id_i}R^{c_1d_1}_{a_1b_1}\ldots R^{c_id_i}_{a_ib_i}
\end{equation}
The symbol $\delta$ denotes a totally antisymmetric product of Kronecker deltas defined as
\begin{equation}\label{Z2}
\delta^{a_1 \ldots a_p}_{b_1 \ldots b_p}=p! \delta^{a_1}_{[b_1} \ldots \delta^{a_p}_{b_p]}
\end{equation}
Here the generalized Kronecker delta symbol is totally antisymmetric in both sets of indices.

The action for Lanczos-Lovelock gravity is given by
\begin{equation}\label{C1}
S=\frac{1}{k_D}\int d^Dx\sqrt{-g} \sum_{i=0}^{[\frac{D}{2}]}c_iL_i+S_{matter}
\end{equation}
Following from the variation of this action, the gravitational equation is written in a compact form:
\begin{equation}
\mathcal{G}_{ab}=\sum_{i=0}^{[\frac{D}{2}]}c_iG^i_{ab}=k_D^2T_{ab}
\end{equation}
where
\begin{equation}
G^{(i)f}_e=-\frac{1}{2^{i+1}}\delta^{f a_1b_1 \ldots a_ib_i}_{ec_1d_1 \ldots c_id_i}R^{c_1d_1}_{a_1b_1}\ldots R^{c_id_i}_{a_ib_i}
\end{equation}
is deduced from the variation of $L_i$ and $T_{ab}$ is the energy-momentum tensor for matter fields obtained from the matter action $S_{matter}$.

$L_0$ is set to 1 so that the constant $c_0$ is proportional to the cosmological constant. $L_1$ gives the usual curvature scalar term. the constant $c_1$ must be positive in order that the Einstein's general relativity in the low energy limit be recovered. $L_2$ is the Gauss-Bonnet term. Although the lagrangian L consists of some higher derivative curvature terms, the Lanczos-Lovelock gravity is not essentially a higher derivative gravity since its equations of motion do not contain terms with more than second derivatives of metric. This reason makes the Lanczos-Lovelock gravity free of ghost \cite{a7}.

\section{Universal thermodynamics: Basics}

Let us consider a (n+1) dimensional homogeneous and isotropic FRW metric as
%%%%%%%%%%%%%%%%%%%%%%%%%%%%%%%%%%%%%%%%%%%%%%%%%%%%%%%%%%%%%
\begin{eqnarray}
ds^2&=&-dt^2 + \frac{a^2(t)}{1-kr^2}dr^2 + R^2 d\Omega_{n-1}^2
\nonumber
\\
&=&h_{ab}dx^a dx^b + R^2 d\Omega_{n-1}^2,
\end{eqnarray}

where $R=ar$ is the area radius, $h_{ab}=diag(-1, \frac{a^2(t)}{1-kr^2})$ is the metric of 2 space ($x^0 =t,~x^1=r$) and $k=0,~\pm1$ denotes flat, open or closed Universe.

The radius of the event horizon ($R_E$) and that of the apparent horizon ($R_A$) can be expressed as

\begin{equation}\label{1}
R_E=a\int_a^\infty \frac{da}{a^2 H}
\end{equation}

\begin{equation}\label{2}
\textrm{and}~R_A=\frac{1}{\sqrt{H^2+\frac{k}{a^2}}}
\end{equation}

respectively, where $H(=\frac{\dot{a}}{a})$ is the Hubble parameter.

The surface gravity ($\kappa$) \cite{r10} on any horizon is given by
\begin{equation}
\kappa=\frac{1}{2\sqrt{-h}}\partial_a(\sqrt{-h}h^{ab}\partial_b R),
\end{equation}
which can be evaluated to give
\begin{equation}
 \kappa = -\left(\frac{R_h}{R_A}\right)^2\left(\frac{1-\frac{\dot{R_A}}{2HR_A}}{R_h}\right),
\end{equation}
%%%%%%%%%%%%%%%%%%%%
i.e.,
%%%%%%%%%%%%%%%%%%%%
\begin{eqnarray}
 \kappa_h&=&-\left(\frac{R_h}{R_A^2}\right)(1-\epsilon), \textrm{on any horizon,}
\nonumber
\\
\textrm{and}~~\kappa_A&=&-\frac{(1-\epsilon)}{R_A}, \textrm{on the apparent horizon,}
\end{eqnarray}
%%%%%%%%%%%%%%%%%%
with $\epsilon= \frac{\dot{R_A}}{2HR_A}$.

Using this form of surface gravity, the extended Hawking temperature is defined as \cite{r11.1,r11.2},
\begin{equation}
T_{EH}^h=\frac{|\kappa_h|}{2\pi }.
\end{equation}
Now the unified first law (UFL) can be expressed as \cite{r7.1,r7.2,r7.3},
\begin{equation}
dE=A\psi +WdV
\end{equation}
where
\begin{equation}
E=\frac{n(n-1)\Omega_nR_h^{n-2}}{16\pi G}(1-h^{ab}\partial_aR\partial_bR)|_{R=R_h},
\end{equation}
is the total energy inside a sphere of radius $R_h$ and is termed as Misner-Sharp energy \cite{r10} (here $\Omega_n=\frac{\pi^{\frac{n}{2}}}{\Gamma (\frac{n}{2}+1)}$). A (=$n\Omega_nR_h^{n-1}$) is the area of the horizon and V (=$\Omega_n R_h^n$) is the volume enclosed by the horizon. The energy flux $\psi$ is termed as the energy supply vector and W is the work function. They are defined as
\begin{eqnarray}
\psi_a&=&T^b_a\partial_br+W\partial_ar,
\nonumber
\\
W&=&-\frac{1}{2}trace T,
\end{eqnarray}
respectively, where $T_{ab}$ is the energy-momentum tensor.

The validity of GSLT and TE can be examined by the following inequalities \cite{r12,r13}
\begin{eqnarray}\label{in}
\frac{d S_{TH}}{d t}& \geq & 0~(\textrm{for GSLT}),
\nonumber
\\
\textrm{and}~\frac{d^2 S_{TH}}{d t^2}& < & 0~(\textrm{for TE}).
\end{eqnarray}
where $S_{TH}=S_h+S_{fh}$, with $S_h$ and $S_{fh}$ as the  horizon entropy and the entropy of the fluid bounded by the horizon respectively. $S_{fh}$ can be calculated from Gibb's relation \cite{r14.1,r14.2,r14.3,r14.4,r14.5,r14.6,r15}
\begin{equation}\label{G}
T_fdS_{fh}=dE_h+pdV_h,
\end{equation}
where $E_h$ is the energy flow across the horizon, $V_h$ is the volume of the fluid and $T_f$ is the temperature of the fluid. 

In the present work, we assume the temperature $T_f$ of the cosmic fluid (chosen as perfect fluid with constant equation of state, modified Chaplygin gas and holographic dark energy) inside the horizon to be same as the extended Hawking temperature ($T_{EH}^h$) of the bounding horizon. But in general, these two temperatures may not be identical. Then there is a spontaneous flow of energy between the horizon and the fluid and is not consistent with FRW model \cite{r15,d3}. So it is natural to assume $T_f \varpropto T_{EH}^h$ \cite{d4} i.e., $T_f=bT_{EH}^h$, with b as the real proportionality constant. It is to be noted that at present, the horizon temperature (i.e., the extended Hawking temperature) is lower than the CMB temperature by many orders of magnitude, so 'b' is restricted as $b\nless 1$. However, in the perspective of local equilibrium hypothesis, there is no spontaneous flow of energy between the horizon and the fluid, so it is reasonable to assume $b=1$. Further due to mathematical complexity of non-equilibrium thermodynamical prescription, the assumption of equilibrium (though restrictive) is widely used in the literature \cite{r14.1,r14.2,r14.3,r14.4,r14.5,r14.6,r15,k1,k2,k3,k4,k5,k6,k7}. Moreover, it should be mentioned that this restrictive situation is suitable at the late stages of the Universe when the cosmic fluid and the horizon will interact for a long time \cite{d5,d6}.  

Now from $(\ref{G})$, the time variation of the entropy of the fluid is given by
\begin{equation}
\frac{dS_{fh}}{dt}=\frac{n\Omega_nR_h^{n-1}}{T^h_{EH}}(\rho +p)\left(\dot{R}_h-HR_h\right).
\end{equation}

\section{Thermodynamics study in Lanczos-Lovelock Gravity Theory}

The Friedmann equation of  FRW Universe in Lanczos-Lovelock gravity is \cite{r10} (assuming G=1)
\begin{equation}\label{f}
\sum_{i=1}^m \hat{c_i}\left(H^2+\frac{k}{a^2}\right)^i=\frac{16\pi \rho}{n(n-1)}
\end{equation}
where $m=[\frac{n+1}{2}]$ and
\begin{eqnarray}
\hat{c_0}&=&\frac{c_0}{n(n-1)},
\nonumber
\\
\hat{c_1}&=&1,
\nonumber
\\
\hat{c_i}&=&c_i \prod_{j=3}^{2m} (n+1-j),~\textrm{for $i>1$}.
\end{eqnarray}

From ($\ref{f}$) we have
\begin{eqnarray}\label{A}
H^2+\frac{k}{a^2}&=&\frac{16 \pi \rho_t}{n(n-1)},
\nonumber
\\
\dot{H}-\frac{k}{a^2}&=&-\frac{8\pi}{n-1}(\rho_t+p_t),
\end{eqnarray}
where $\rho_t=\rho+\rho_e$, $\rho_t+p_t=(\rho+p)+(\rho_e+p_e)$ and
\begin{eqnarray}\label{B}
\rho_e&=&-\frac{n(n-1)}{16\pi}\sum_{i=2}^m \frac{ \hat{c_i}}{R_A^{2i}},
\nonumber
\\
\rho_e+p_e&=&-\frac{\epsilon (n-1)}{4 \pi}\sum_{i=2}^m\frac{ i\hat{c_i}}{R_A^{2i}}.
\end{eqnarray}
For this modified gravity theory, we shall determine the expression for horizon entropy from the validity of the unified first law of thermodynamics both for an event and an apparent horizon.
\subsection{Expressions of Entropy}
For the present model the energy flux vector can be written as
\begin{equation}
\psi=\psi_m+\psi_e,
\end{equation}
where
\begin{eqnarray}\label{C}
\psi_m&=&-\frac{1}{2}(\rho+p)HRdt+\frac{1}{2}(\rho+p)a dr,
\nonumber
\\
\psi_e&=&-\frac{1}{2}(\rho_e+p_e)HRdt+\frac{1}{2}(\rho_e+p_e)a dr.
\end{eqnarray}

We consider the tangent vector to the surface of event horizon as \cite{r11.1,r11.2}
\begin{equation}
\xi_E=\partial_t-\frac{1}{a}\partial_r.
\end{equation}

Now projecting the UFL along $\xi_E$, the first law of thermodynamics of the event horizon is obtained as \cite{r9,r10,r11.1,r11.2}
\begin{equation}\label{M}
\langle dE, \xi_E \rangle =\frac{\kappa_E}{8\pi}\langle dA, \xi_E \rangle +\langle WdV, \xi_E \rangle.
\end{equation}
The matter energy supply $A\psi_m$, when projected on the event horizon gives the heat flow $\delta Q$ in the Clausius relation ($\delta Q=TdS$). So from $(\ref{M})$, we have
\begin{equation}\label{N}
dQ=\langle A\psi_m, \xi_E \rangle =\frac{\kappa_E}{8\pi}\langle dA, \xi_E. \rangle-\langle A\psi_e, \xi_E \rangle
\end{equation}

Now using equations ($\ref{A} ),~( \ref{B}),~ (\ref{C}$), and the extended Hawking temperature on the event horizon and comparing with the Clausius relation and integrating, the entropy on the event horizon can be evaluated as

\begin{equation}\label{ehe}
S_E=\frac{A}{4}+\frac{n\Omega_n (n-1)}{4}\int \left(\frac{\epsilon R_E^{n-2}R_A^2}{1-\epsilon}\right)\left(\frac{HR_E+1}{HR_E-1}\right)\left(\sum_{i=2}^m\frac{i\hat{c_i}}{R_A^{2i}}\right)dR_E.
\end{equation}
 Similarly for the apparent horizon, considering \cite{r9}
\begin{equation}\label{ah}
\xi_A=\partial_t-(1-2\epsilon)Hr\partial_r
\end{equation}
as the tangent vector to the surface of the apparent horizon, the expression for the  entropy becomes,
\begin{equation}\label{ahe}
S_A=\frac{A}{4}+\frac{n\Omega_n (n-1)}{2}\int HR_A^{n+1}\epsilon \left(\sum_{i=2}^{m}\frac{i\hat{c_i}}{R_A^{2i}}\right)dt.
\end{equation}

From equations (\ref{ehe}) and (\ref{ahe}) we see that the entropy at the horizon (event/ apparent) is no longer the usual Bekenstein entropy, rather there are correction terms with the Bekenstein's area formula. Due to complicated expressions, the correction terms are only presented in integral form.
\subsection{Thermodynamical analysis}

Now the time derivative of the total entropy (i.e., entropy of the horizon+ entropy of the fluid) are given by (for simplicity flat space-time has been considered),
\begin{equation}
\dot{S}_{TE}=\frac{n(n-1)\Omega_n}{4}\left[v_ER_E^{n-2}+\frac{v_AR_E^{n-2}(v_E+2)}{2-v_A}\left(\sum_{i=2}^m \frac{i\hat{c_i}}{R_A^{2(i-1)}}\right)\right]-\frac{n(n-1)\Omega_nR_E^{n-2}v_A}{2(2-v_A)}.
\end{equation}
\begin{equation}
\dot{S}_{TA}=\frac{n\Omega_n(n-1)}{4}\left[R_A^{n-2}v_A+R_A^nv_A\left(\sum_{i=2}^m\frac{i\hat{c_i}}{R_A^{2i}}\right)\right]+\frac{n(n-1)\Omega_n R_A^{n-2}v_A(v_A-1)}{2(2-v_A)}.
\end{equation}
where $v_A~(=\dot{R}_A)$ and $v_E~(=\dot{R}_E)$ are the velocities of the apparent and event horizon respectively and $S_{TA}=S_A+S_{fA}$ and $S_{TE}=S_E+S_{fE}$.\\
Again taking the time derivative of $\dot{S}_{TE}$ and $\dot{S}_{TA}$ we have
\begin{eqnarray}
\ddot{S}_{TE}&=&\frac{n(n-1)\Omega_n}{4}\left[(n-2)R_E^{n-3}v_E^2+f_ER_E^{n-2}+\frac{v_AR_E^{n-2}(v_E+2)}{2-v_A}\{\sum_{i=2}^m\frac{i\hat{c_i}}{R_A^{2(i-1)}}\}\left(\frac{n-2}{R_E}v_E+\frac{f_E}{v_E+2}+\frac{2f_A}{v_A(2-v_A)}\right)\right.
\nonumber
\\
&-&\left.\frac{v_A^2R_E^{n-2}(v_E+2)}{2-v_A}\left(\sum_{i=2}^m\frac{2i(i-1)\hat{c_i}}{R_A^{2i-1}}\right)\right]
-\frac{n(n-1)\Omega_n}{2}\left[\frac{R_E^{n-2}v_A}{2-v_A}\{\frac{n-2}{R_E}v_E+\frac{f_A}{v_A}+\frac{f_A}{2-v_A}\}\right]
\end{eqnarray}
\begin{eqnarray}
\ddot{S}_{TA}&=&\frac{n(n-1)\Omega_n}{4}\left[R_A^{n-3}\{(n-2)v_A^2+R_Af_A\}+\left(\sum_{i=2}^m\frac{i\hat{c_i}}{R_A^{2i}}\right)R_A^nv_A^2\left(\frac{nv_A}{R_A}+\frac{f_A}{v_A}\right)-R_A^nv_A^2\left(\sum_{i=2}^m\frac{2i^2\hat{c_i}}{R_A^{2i+1}}\right)\right]
\nonumber
\\
&+&\frac{n(n-1)\Omega_n}{2}\left[\frac{R_A^{n-2}v_A(v_a-1)}{2-v_a}\left(\frac{n-2}{R_A}v_A+f_A\{\frac{1}{v_A}+\frac{1}{v_A-1}+\frac{1}{2-v_A}\}\right)\right]
\end{eqnarray}
where $f_E~(=\dot{v}_E)$ and $f_A~(=\dot{v}_A)$ are the acceleration of the event and apparent horizon respectively.\\\\

{\bf $\bullet$ Perfect fluid (with constant equation of state)}\\\\
When the Universe is considered to be filled by a perfect fluid having a constant equation of state,
\begin{equation}
p=\omega \rho,
\end{equation}
then the velocity of the apparent and event horizons are evaluated as
\begin{eqnarray}\label{OP}
v_A&=& \dot{R}_A=-\frac{\dot{H}}{H^2},
\nonumber
\\
and~v_E&=&\dot{R}_E=HR_E-1
\end{eqnarray}
respectively.\\
Due to complicated expression, the expressions for $\dot{S}_{TA}$ and $\dot{S}_{TE}$ (using equation (\ref{OP})) have been represented in Figs. 1(a)-(d) against the equation of state parameter $\omega$ for the following choices of the different parameters involved: $H=1,~R_E=3,~c_0=1,~c_1=1,~c_2=2,~c_3=3,~c_4=4$. The constraints on $\omega$ for which the GSLT and TE holds in the case of both the apparent and event horizons have been presented in tabular form of Table 1.
\begin{figure}
\begin{minipage}{0.4\textwidth}
\includegraphics[width=1.0\linewidth]{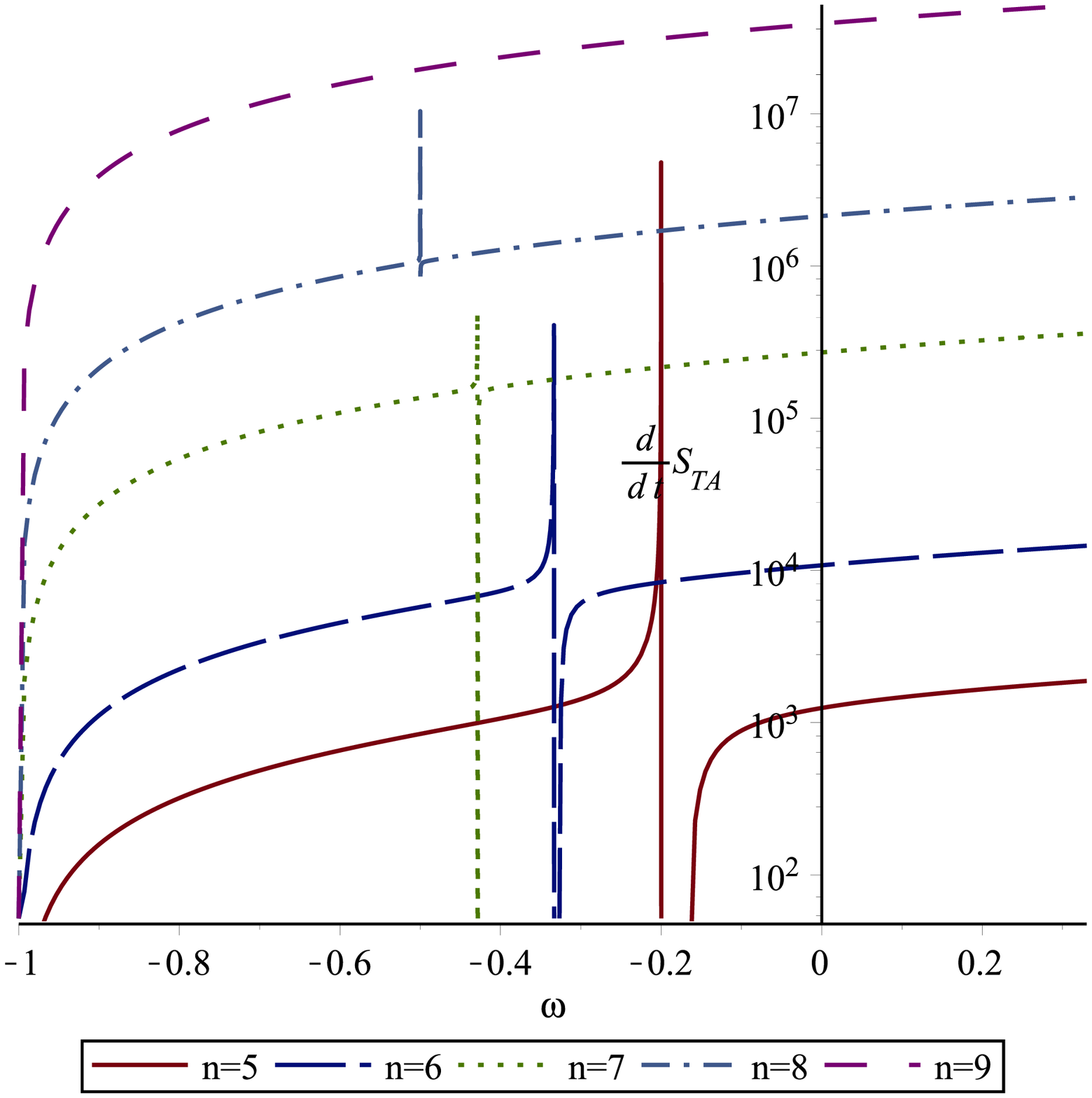}
(a): GSLT when the Universe is bounded by the apparent horizon.%\caption{}
\end{minipage}
\begin{minipage}{0.4\textwidth}
\includegraphics[width=1.0\linewidth]{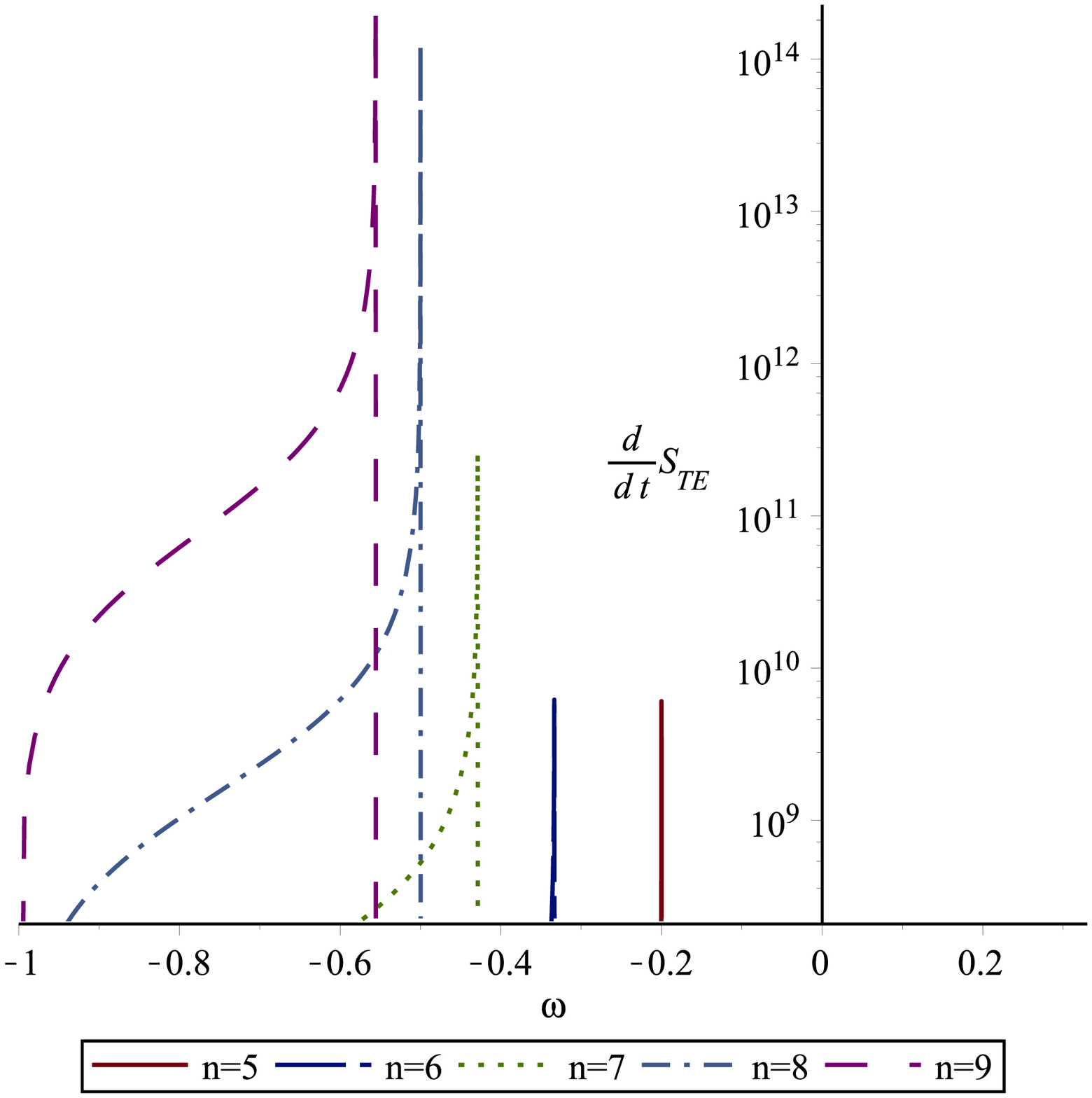}
(b): GSLT when the Universe is bounded by the event horizon%\caption{}
\end{minipage}
\begin{minipage}{0.4\textwidth}
\includegraphics[width=1.0\linewidth]{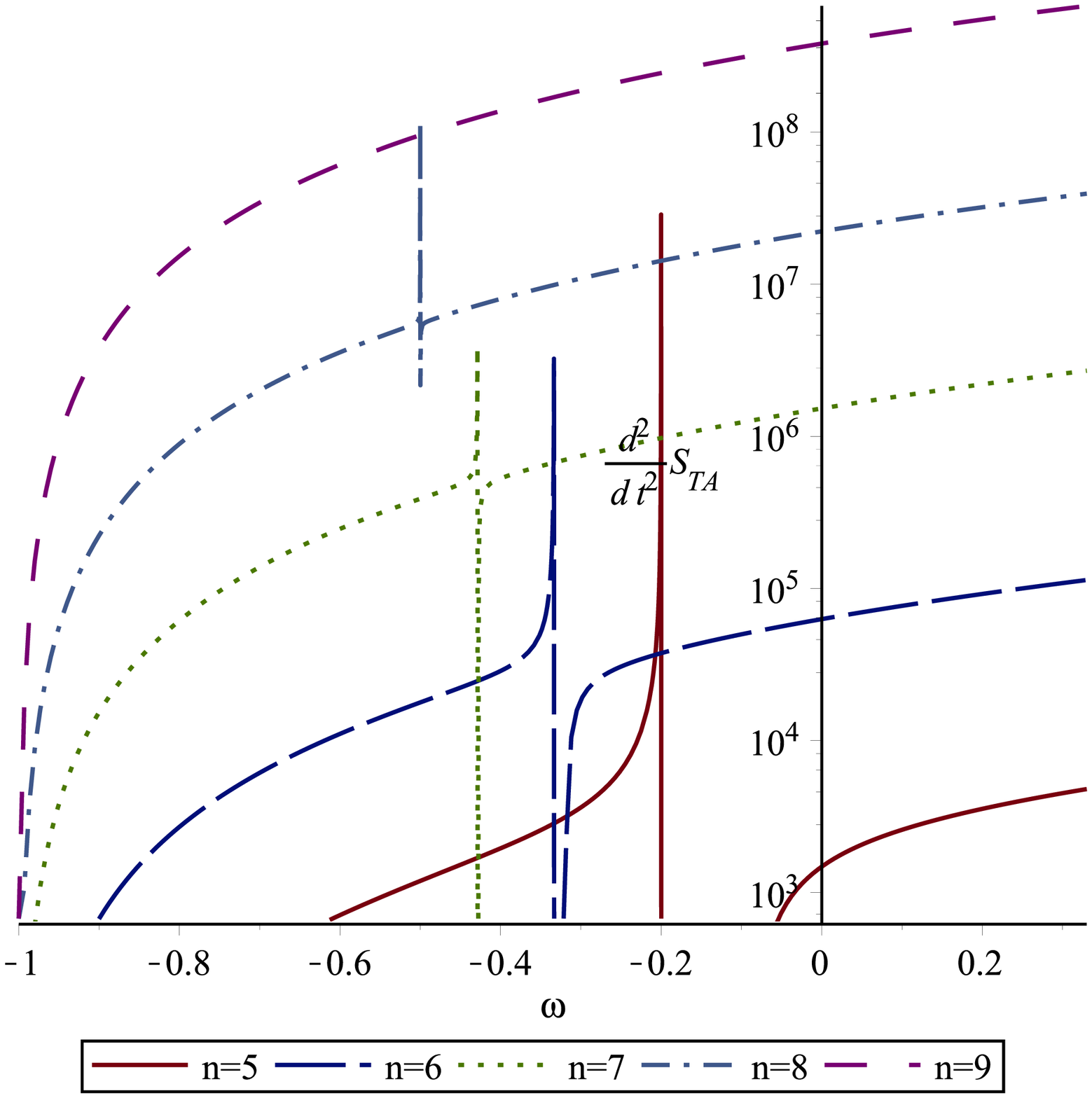}
(c) TE when the Universe is bounded by the apparent horizon %\caption{}
\end{minipage}
\begin{minipage}{0.4\textwidth}
\includegraphics[width=1.0\linewidth]{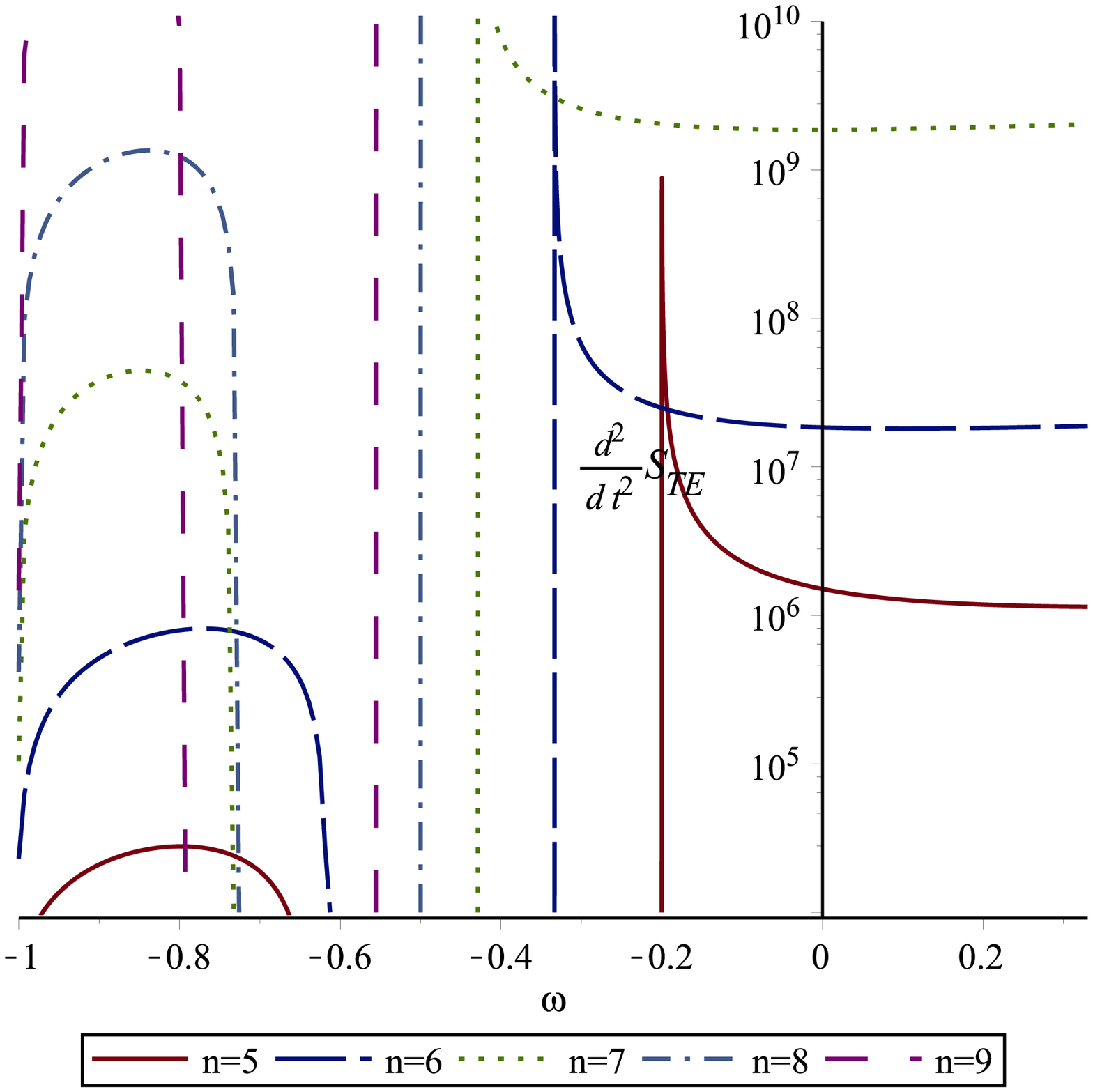}
(d): TE when the Universe is bounded by the event horizon%\caption{}
\end{minipage}
\caption{The plots shows GSLT and TE for universe filled with a perfect fluid having a constant equation of state }
\end{figure}

\begin{center}
Table 1: Constraints on $\omega$ for which the GSLT and the TE holds for a perfect fluid having a constant equation of state and bounded by event horizon (EH) or apparent horizon (AH)
\end{center}
\begin{center}
\begin{tabular}{|c|c|c|c|c|}
\hline Dimension & GSLT( EH)&GSLT(AH) & TE (EH)& TE(AH)\\
\hline \hline 6& $\omega < -0.2$ & $\omega <-0.2~or ~-0.175 \lesssim \omega$ & $\omega < -0.2$ & $-0.2 <\omega \lesssim -0.1$\\
\hline 7&$\omega \lesssim -0.34$&$\omega \lesssim-0.33~or~-0.32\lesssim \omega$&$\omega \lesssim -0.34$&$-0.335 \lesssim \omega \lesssim -0.31$\\
\hline 8& $\omega \lesssim -0.43$ & Always hold & $\omega \lesssim -0.428$ &Never  hold\\
\hline 9& $\omega \lesssim -0.5$& Always hold &$\omega \lesssim -0.5$& Never hold\\
\hline 10& $\omega \lesssim -0.55$ & Always hold & $\omega \lesssim -0.55$ & Never  hold\\
\hline
\end{tabular}
\end{center}

{\bf $\bullet$ Modified Chaplygin Gas}\\\\
The modified Chaplygin gas (MCG) model unifies dark matter and dark energy. At low density, the fluid can accommodate late time acceleration while a radiation dominated era can be obtained at high density. Thus the MCG model can describe the evolution of the Universe starting from the radiation epoch to the epoch dominated by the dark energy. The dynamics of the MCG model was studied by Wu et al. \cite{A} whereas Bedran et al. \cite{B} studied the evolution of the temperature function in the presence of a MCG. It is also consistent with perturbative studies \cite{C} and the spherical collapse problem \cite{D}.

The equation of state for MCG is given by \cite{r13,r19}
\begin{equation}
p=\gamma \rho-\frac{B}{\rho^{n_1}}
\end{equation}
where $\gamma~(\leq 1) $ , $B, ~n_1$ are positive constant. Now solving the energy conservation equation (in (n+1)-dimensions)
\begin{equation}
\dot{\rho}-nH(\rho+p)=0
\end{equation}
we have
\begin{equation}
\rho^{1+n_1}=\frac{1}{\gamma +1}\left[B+\left(\frac{C}{a^n}\right)^{(\gamma +1)(n_1+1)}\right]
\end{equation}
where C is an arbitrary constant of integration. \\
In this model the velocity of the apparent horizon can be evaluated as $$v_A=\frac{nC(1+\gamma)}{2(Ba^\mu +C)},$$ where $\mu=n(1+n_1)(1+\gamma)$. The radius of the event horizon can be expressed in terms of hypergeometric function as $$R_E=R_{1~~2}F_1\left[\frac{1}{2(n_1+1)}, \frac{1}{\mu}, 1+\frac{1}{\mu}, \frac{-C}{Ba^\mu}\right],$$ where $R_1=\frac{\sqrt{\frac{n(n-1)}{16\pi}}(1+\gamma)^{\frac{1}{2(n+1_1)}}}{B^{\frac{1}{2(n_1+1)}}}$.\\
The velocity of the event horizon is $$v_E=HR_E-1.$$
Now we plot $\dot{S}_{TA},~\dot{S}_{TE},~\ddot{S}_{TA}~and~\ddot{S}_{TE}$ against $\gamma$ in Figure:2 (a-d) considering $H=1.5,~a=1,~n_1=0.25,~B=2,~C=1,~c_0=1,~c_1=1,~c_2=2,~c_3=3 ~and ~c_4=4$. From the figures, it is evident that GSLT and TE for the Universe filled with MCG always holds in case of both the apparent and the event horizons.
\begin{figure}
\begin{minipage}{0.4\textwidth}
\includegraphics[width=1.0\linewidth]{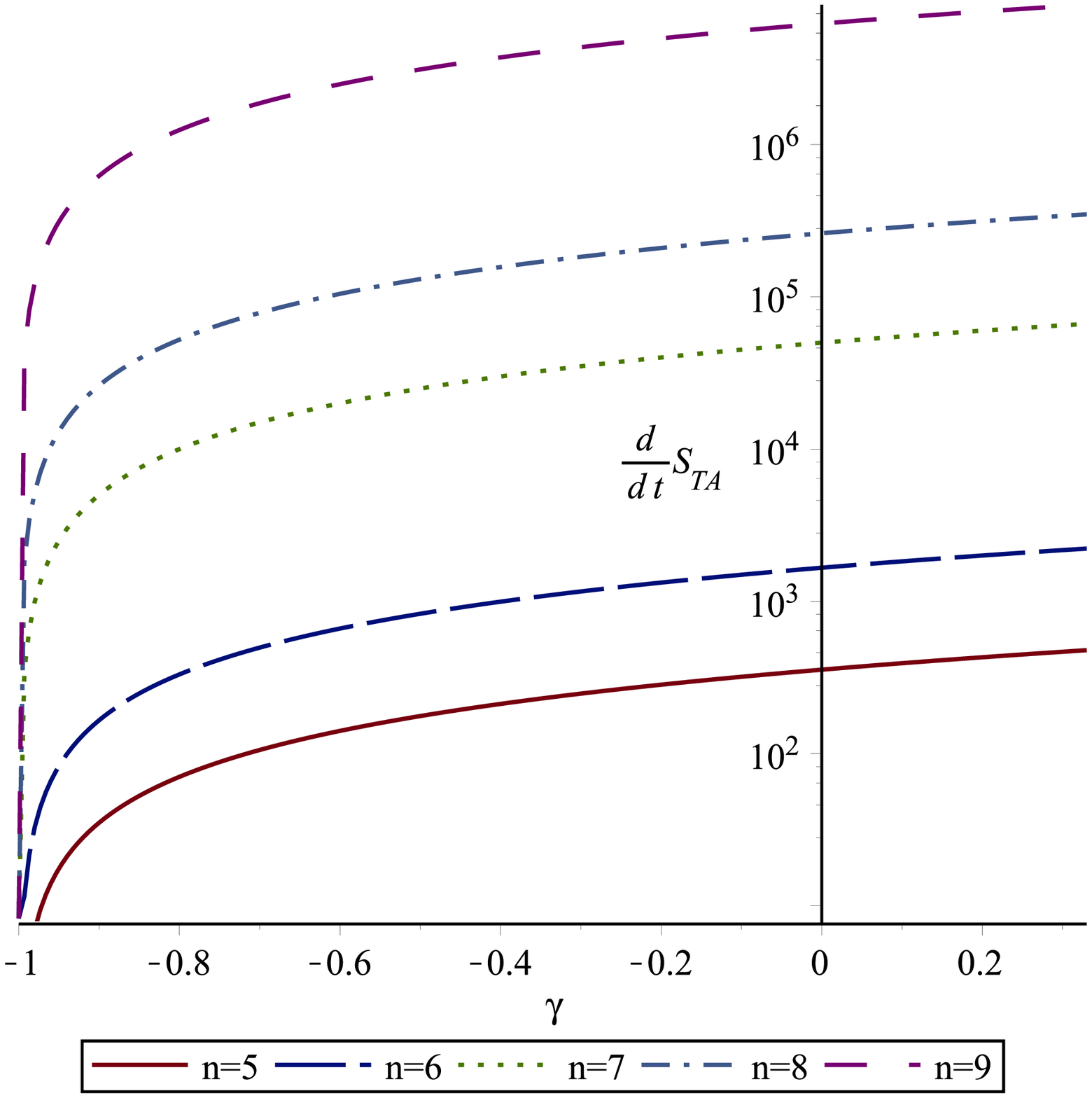}
(a): GSLT when the Universe is bounded by the apparent horizon.%\caption{}
\end{minipage}
\begin{minipage}{0.4\textwidth}
\includegraphics[width=1.0\linewidth]{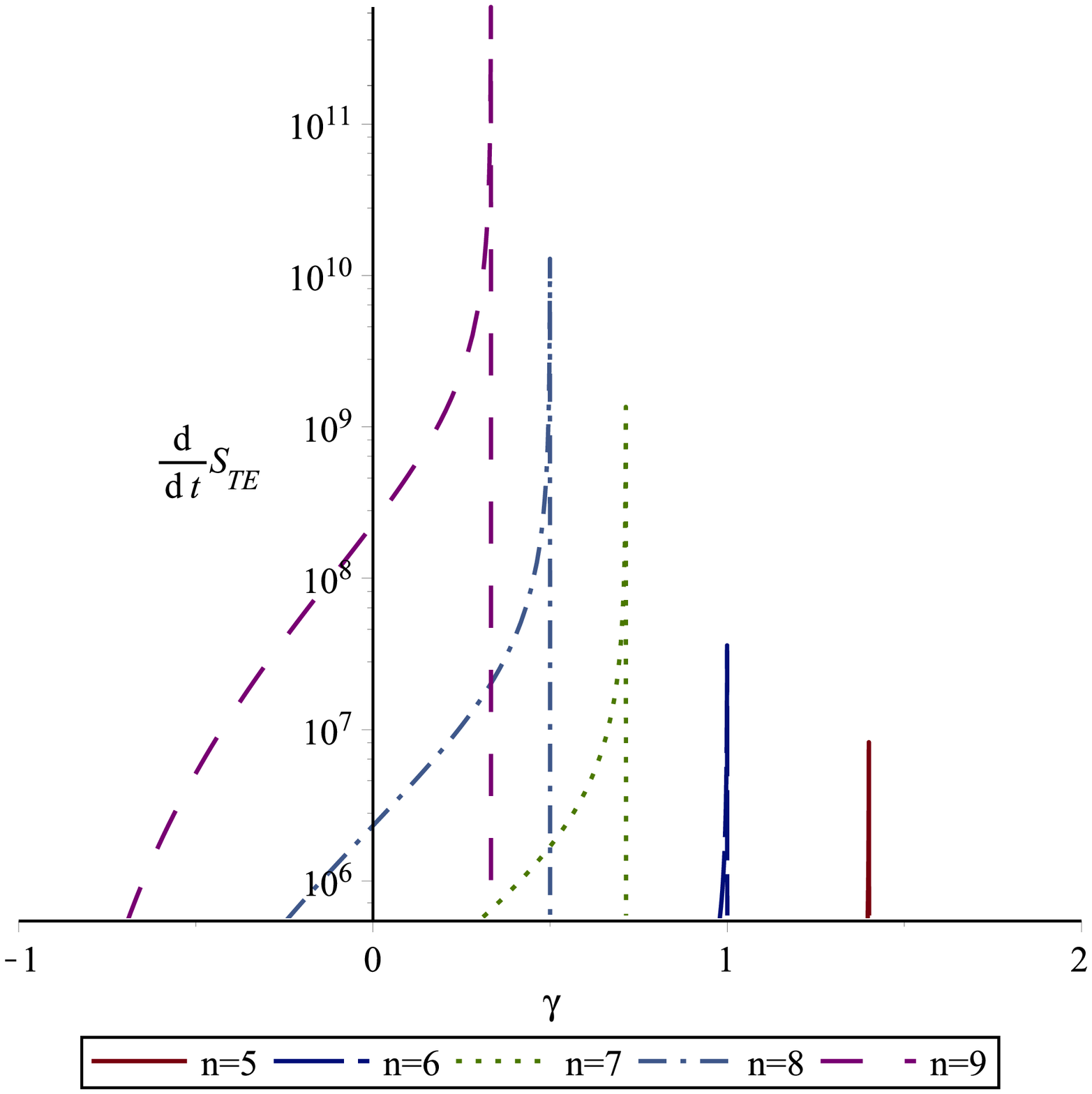}
(b): GSLT when the Universe is bounded by the event horizon%\caption{}
\end{minipage}
\begin{minipage}{0.4\textwidth}
\includegraphics[width=1.0\linewidth]{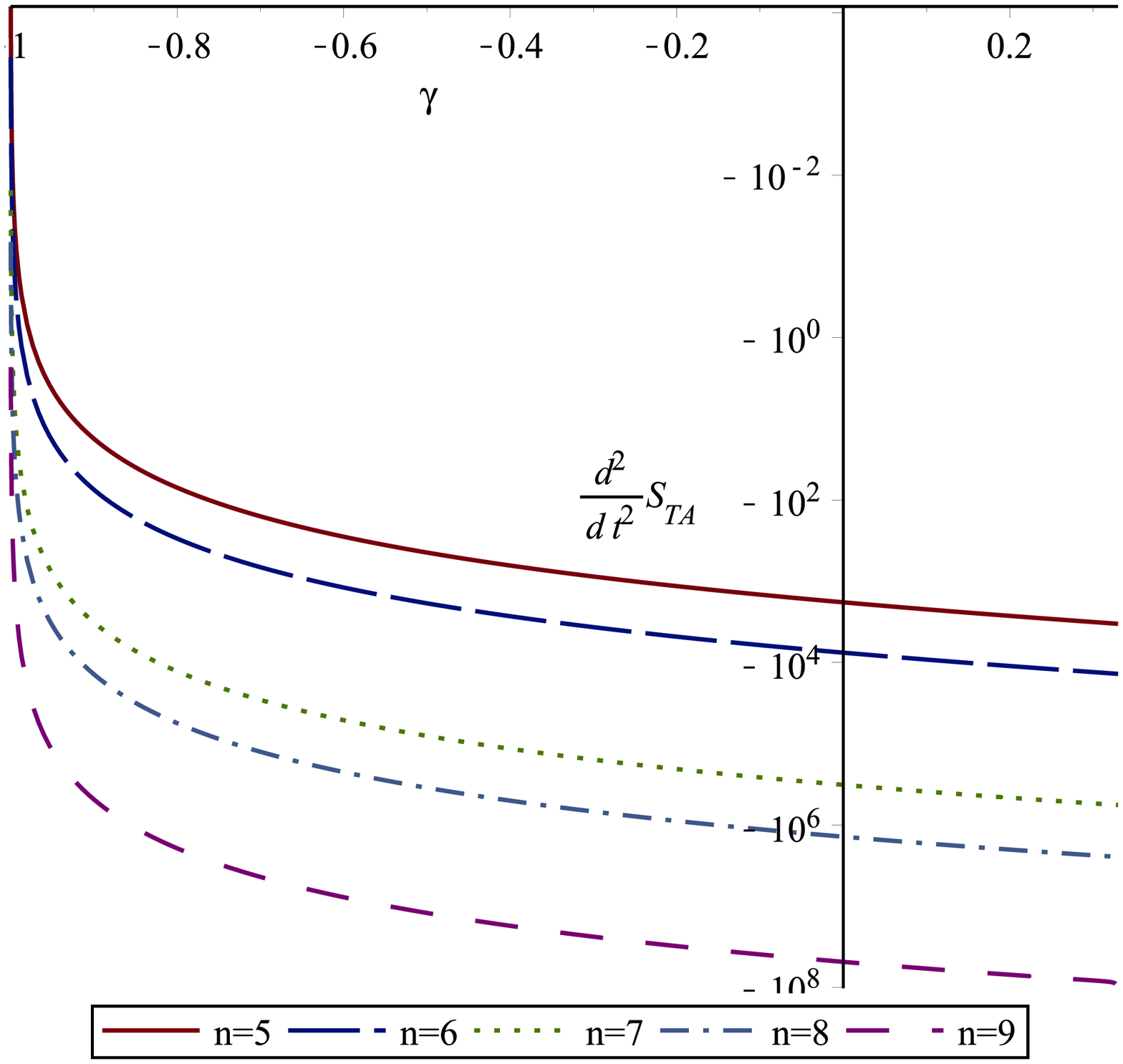}
(c) TE when the Universe is bounded by the apparent horizon %\caption{}
\end{minipage}
\begin{minipage}{0.4\textwidth}
\includegraphics[width=1.0\linewidth]{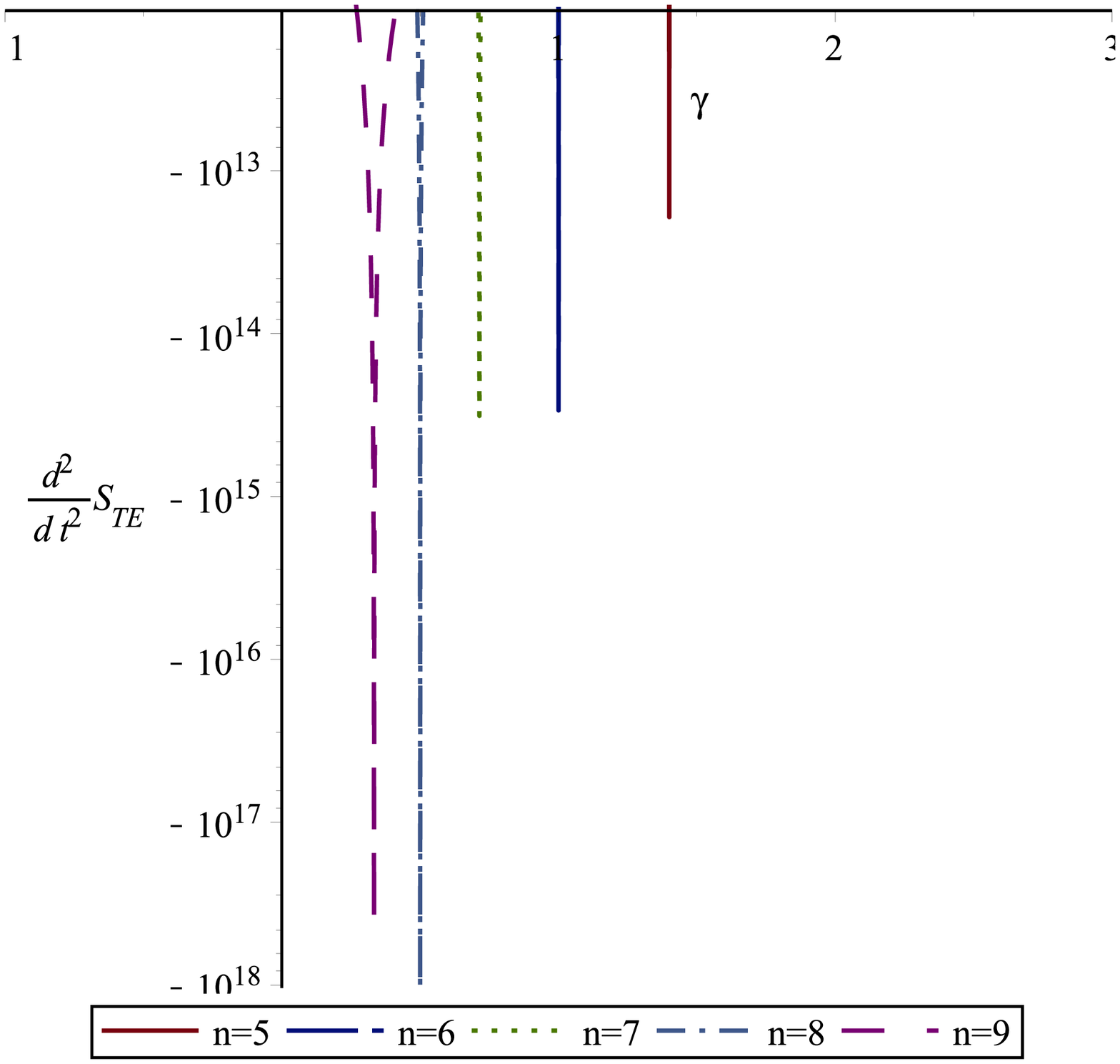}
(d): TE when the Universe is bounded by the event horizon%\caption{}
\end{minipage}
\caption{The plot shows GSLT and TE for Modified Chaplygin Gas model }
\end{figure}

{\bf $\bullet$ Holographic Dark Energy (HDE)}\\\\
The (interacting) HDE models are favoured by observed data obtained from the Cosmic Microwave Background (CMB) \cite{c1} and matter distribution at large scales \cite{c2}. Further, Das et al. \cite{c3} and Amendola et al. \cite{c4} showed that an interaction (between HDE and DM in the dust form) model of the Universe mimics the observationally measured phantom equation of state as compared to noninteracting models, which may predict a non-phantom type of equation of state.

When the Universe is filled by holographic dark energy, the total energy density is $\rho=\rho_m+\rho_D$, where $\rho_m$ and $\rho_D$ are the energy densities of dark matter (DM) and dark energy (DE) respectively. As the interaction between DM and DE has been considered, $\rho_m$ and $\rho_D$ will satisfy the following conservation laws:\\
\begin{eqnarray}
\dot{\rho}_m+nH\rho_m&=&Q,
\nonumber
\\
\dot{\rho}_D+nH(1+\omega_D)\rho_D&=&-Q
\end{eqnarray}
where $\omega_D$ is the equation of state parameter of DE, Q denotes the interaction term which can be taken as $Q=nHb^2\rho$, with $b^2$ the coupling constant.

We can show that (by the process done in \cite{r20}) the equation of state parameter has the form
\begin{equation}
\omega_d=-\frac{1}{n}-\frac{2\sqrt{\Omega_d}}{nc}-\frac{b^2}{\Omega_d},
\end{equation}

where c is a dimensionless parameter (estimated from observations). The density parameter $\Omega_d$ evolves as
\begin{equation}
\Omega_d'=\Omega_d \left[(1-\Omega_d)\left(1+\frac{2\sqrt{\Omega_d}}{c}\right)-nb^2\right]
\end{equation}
where $'=\frac{\partial}{\partial x},~ x=lna$.\\
In this case $p=p_d$ where $p_d$ thermodynamic pressure of holographic dark energy.
The velocities of the apparent ($v_A$) and event horizon ($v_E$) can be expressed as
\begin{equation}
v_A=\frac{n}{2}\left[(1-b^2)-\frac{\Omega_d}{n}\left(1+\frac{2\sqrt{\Omega_d}}{c}\right)\right]
\end{equation}
and
\begin{equation}
v_E=\left(\frac{c}{\sqrt{\Omega_d}}-1\right)
\end{equation}
The validity of GSLT and TE are examined graphically in the Figure 3-5 (considering H=1,$c_0=1,~c_1=1,~c_2=2,~c_3=3~and~c_4=4$). In the figures, the density parameter $\Omega_d$ and the dimensionless parameter c are estimated from three Planck data sets \cite{r21} (see Table-2).

\begin{center}
{\bf Table 2: Planck Data Sets}
\end{center}
\begin{center}
\begin{tabular}{|c|c|c|c|}
\hline Sl. No. & Data Sets & c & $\Omega_d$ \\
\hline \hline 1 & Planck+CMB+SNLS3+lensing & 0.603 & 0.699 \\
\hline 2 & Planck+CMB+Union 2.1+lensing & 0.645 & 0.679 \\
\hline 3 & Planck+CMB+BAO+HST+lensing & 0.495 & 0.745\\
\hline
\end{tabular}
\end{center}

Planck data reduces the error by 30\%-60\% when compared to  Wilkinson Microwave Anisotropy Probe (WMAP)-9 data. The accuracy is further increased if the External Astrophysical data sets (EADS) as well as lensing data is taken into account. Common EADS include the Baryonic Acustic Oscillation (BAO) measurments from 6dFGS+SDSS DR7(R) + BOSS DR9, Hubble constant estimated from the Hubble Space Telescope (HST) and supernova data sets SNLS3 and Union 2.1.\\

\begin{figure}
\begin{minipage}{0.4\textwidth}
\includegraphics[width=1.0\linewidth]{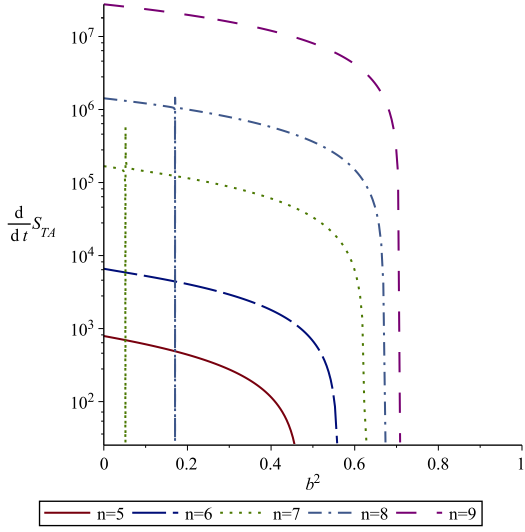}
(a)GSLT wwhen Universe is bounded by the apparent horizon %\caption{}
\end{minipage}
\begin{minipage}{0.4\textwidth}
\includegraphics[width=1.0\linewidth]{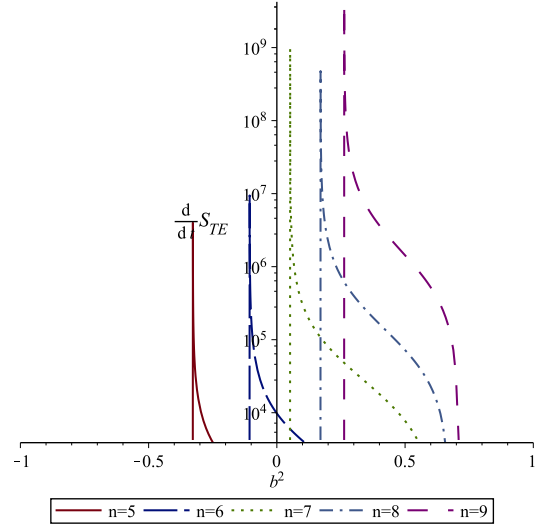}
(b) GSLT when the Universe is bounded by the event horizon%\caption{}
\end{minipage}
\begin{minipage}{0.4\textwidth}
\includegraphics[width=1.0\linewidth]{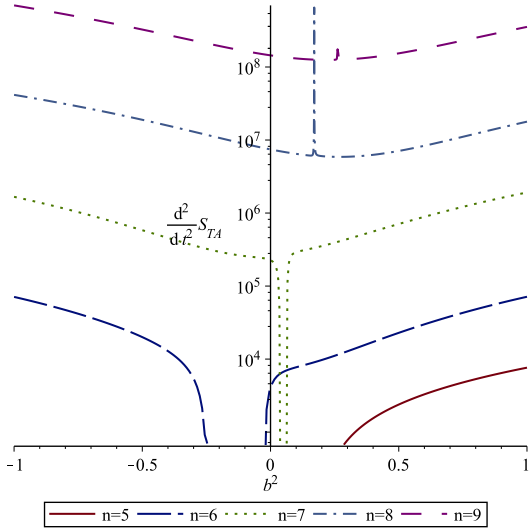}
(c) TE when the Universe is bounded by the apparent horizon%\caption{}
\end{minipage}
\begin{minipage}{0.4\textwidth}
\includegraphics[width=1.0\linewidth]{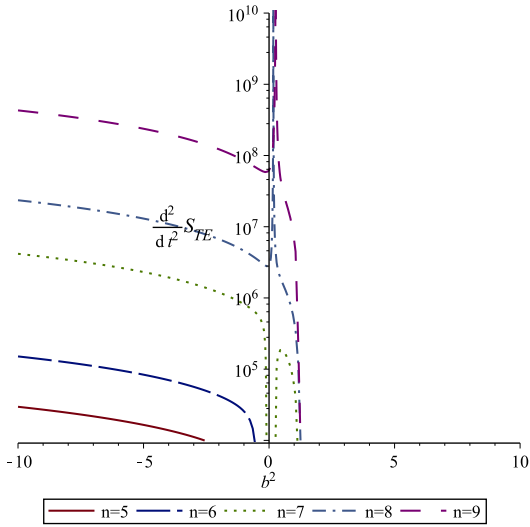}
(d)TE when the Universe is bounded by the event horizon%\caption{}
\end{minipage}
\caption{ The plot, shows the validity of GSLT and TE when Data 1 has been considered for holographic dark energy }
\end{figure}

\begin{figure}
\begin{minipage}{0.4\textwidth}
\includegraphics[width=1.0\linewidth]{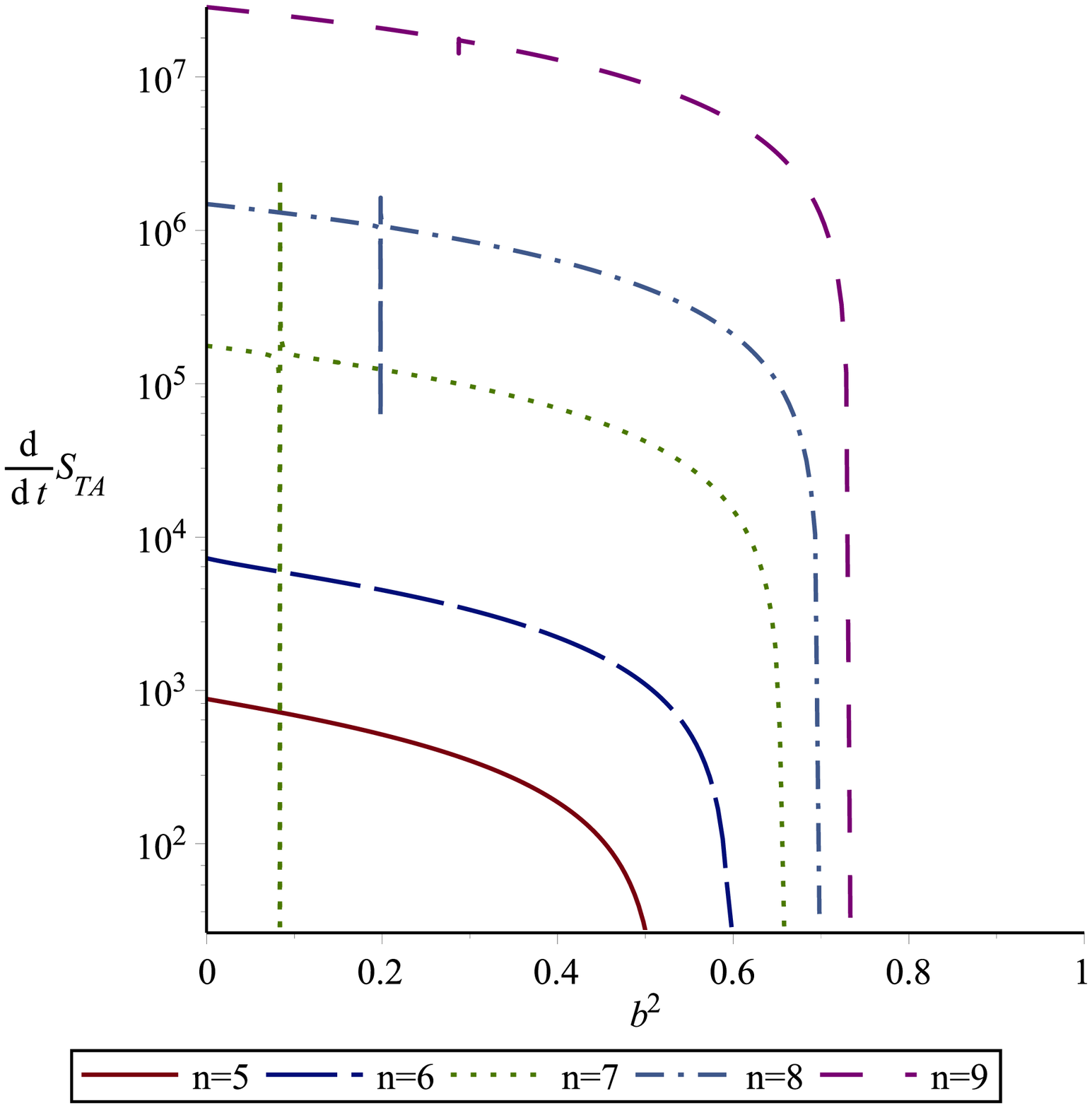}
(a)GSLT wwhen Universe is bounded by the apparent horizon %\caption{}
\end{minipage}
\begin{minipage}{0.4\textwidth}
\includegraphics[width=1.0\linewidth]{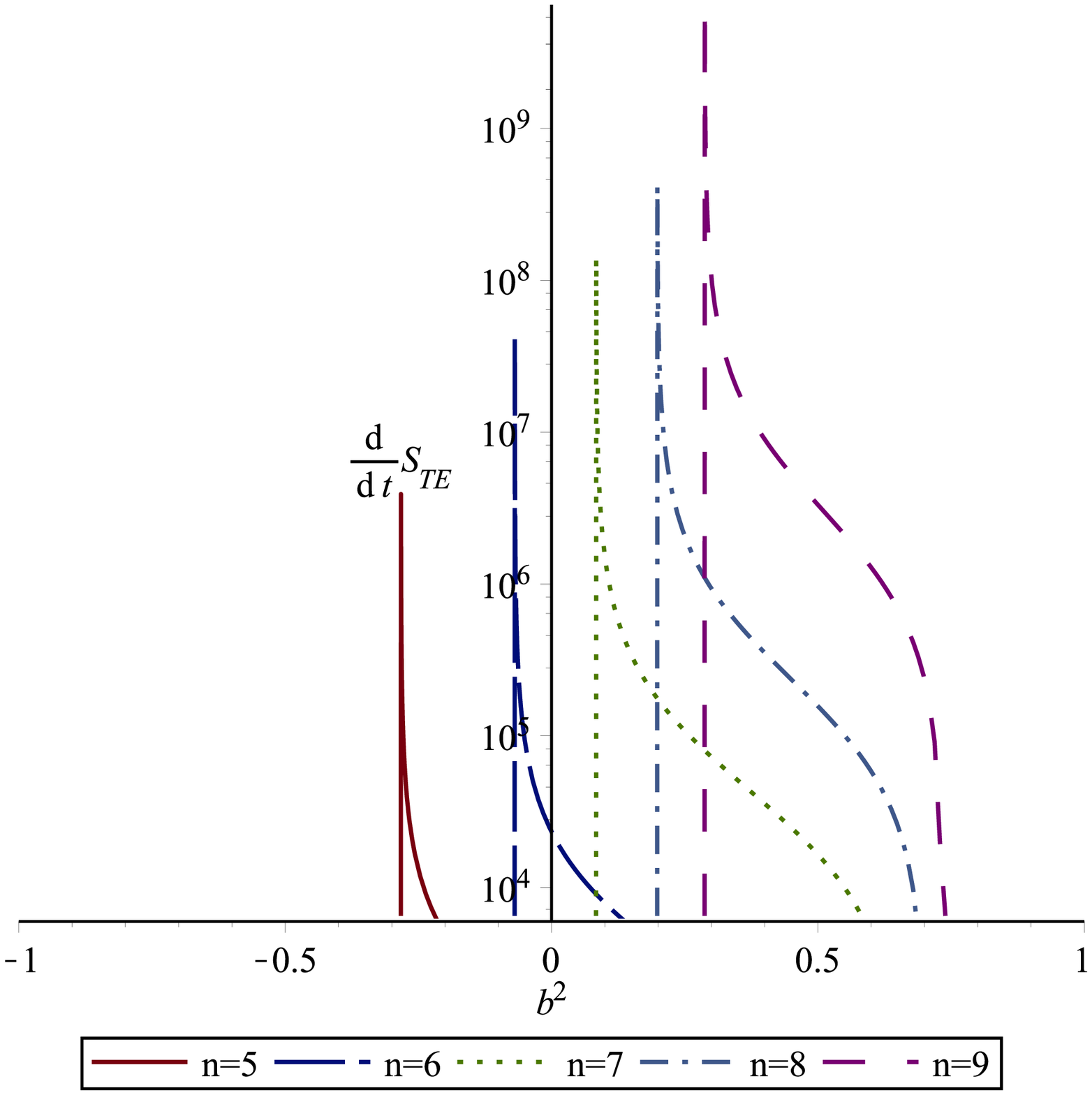}
(b) GSLT when the Universe is bounded by the event horizon%\caption{}
\end{minipage}
\begin{minipage}{0.4\textwidth}
\includegraphics[width=1.0\linewidth]{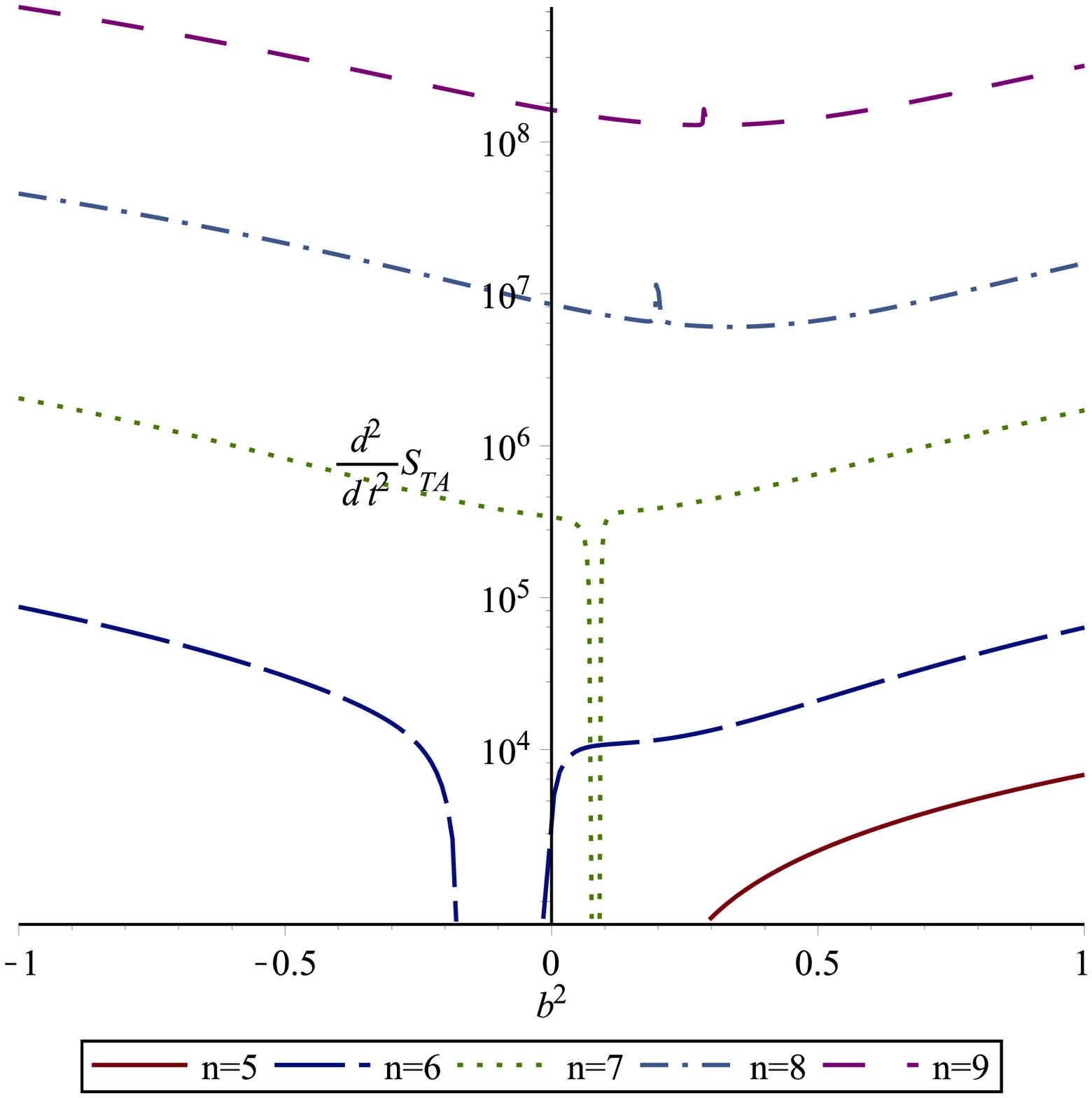}
(c) TE when the Universe is bounded by the apparent horizon%\caption{}
\end{minipage}
\begin{minipage}{0.4\textwidth}
\includegraphics[width=1.0\linewidth]{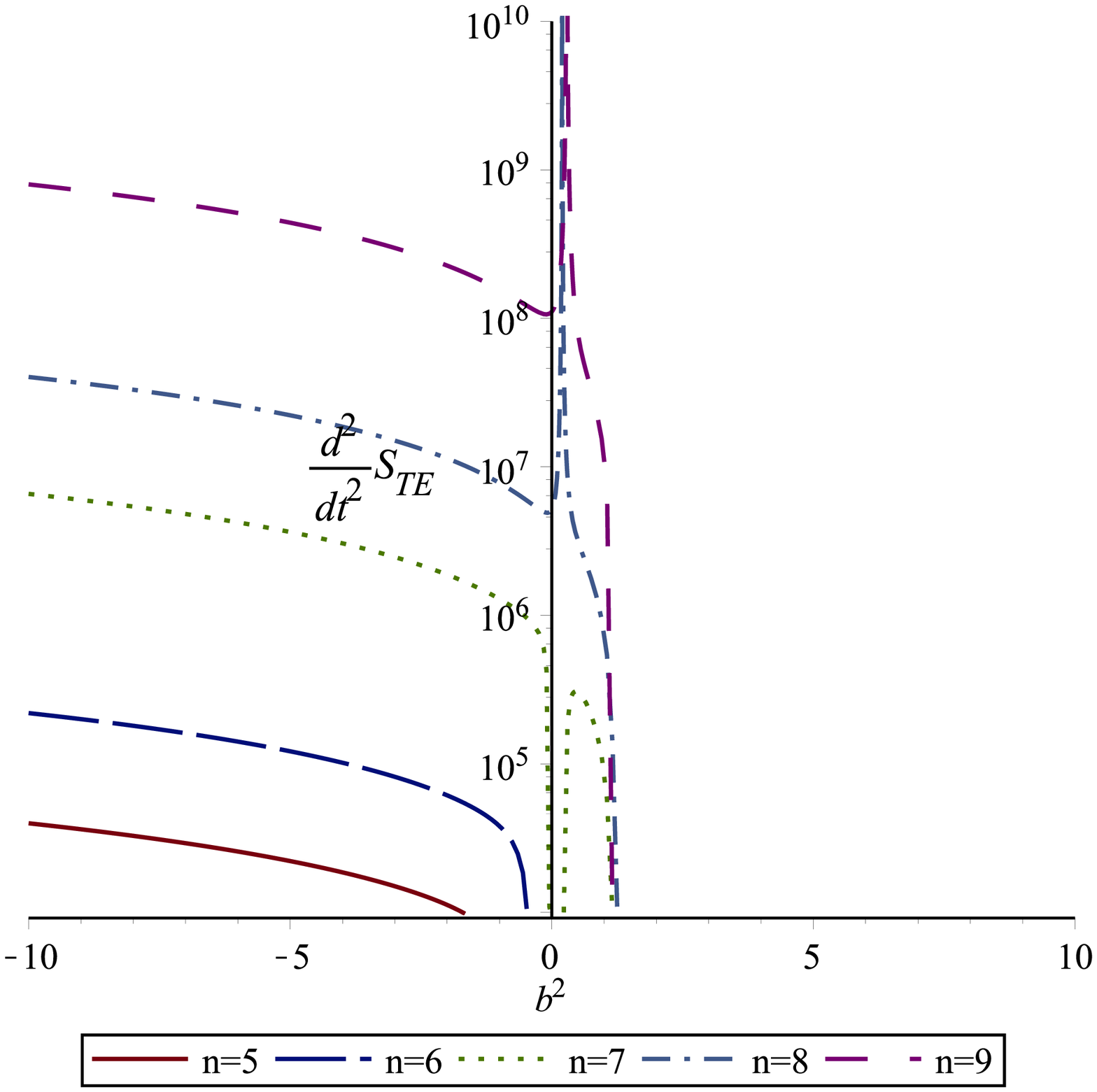}
(d)TE when the Universe is bounded by the event horizon%\caption{}
\end{minipage}
\caption{ The plot, shows the validity of GSLT and TE when Data 2 has been considered for holographic dark energy }
\end{figure}

\begin{figure}
\begin{minipage}{0.4\textwidth}
\includegraphics[width=1.0\linewidth]{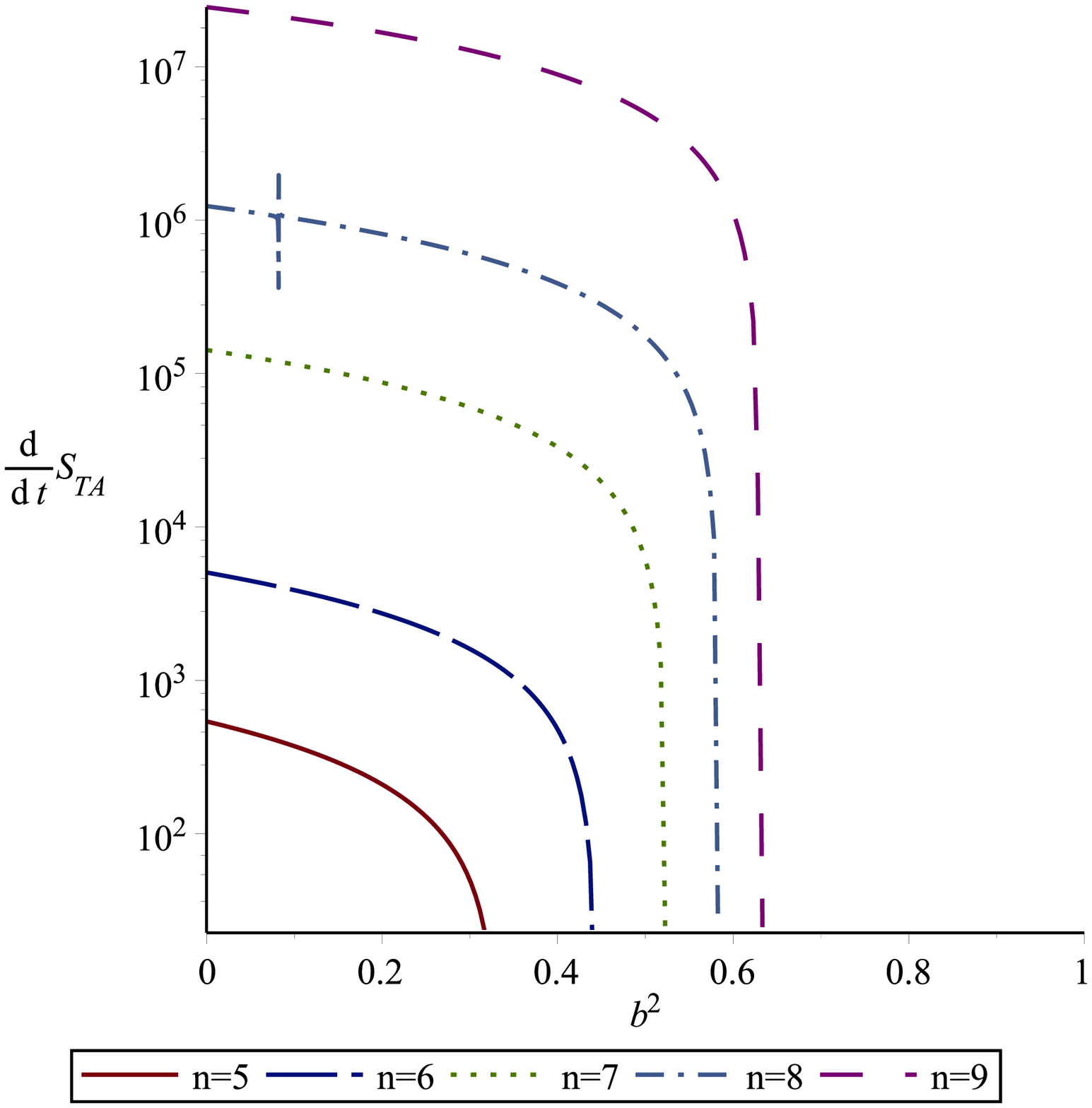}
(a)GSLT wwhen Universe is bounded by the apparent horizon %\caption{}
\end{minipage}
\begin{minipage}{0.4\textwidth}
\includegraphics[width=1.0\linewidth]{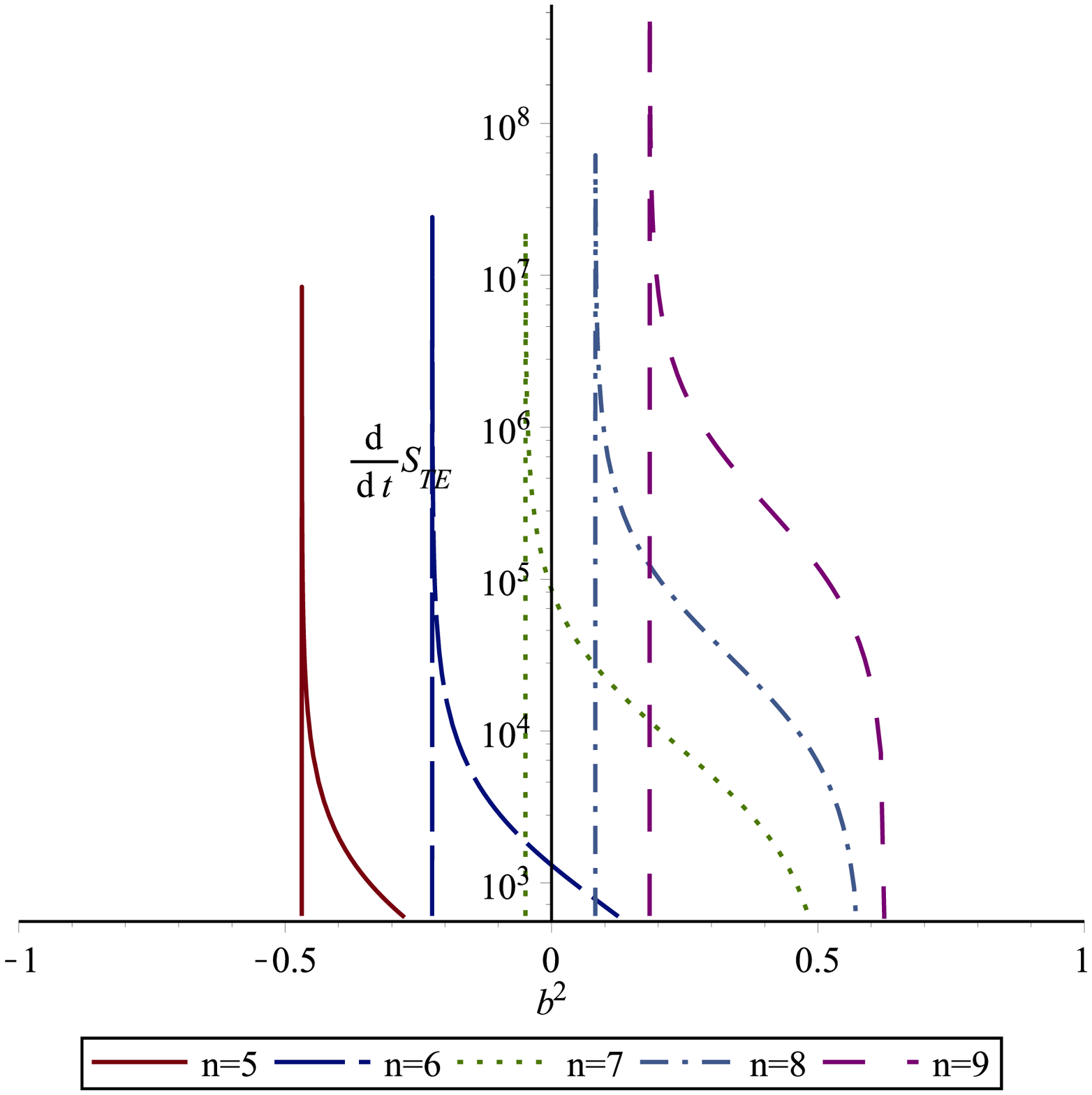}
(b) GSLT when the Universe is bounded by the event horizon%\caption{}
\end{minipage}
\begin{minipage}{0.4\textwidth}
\includegraphics[width=1.0\linewidth]{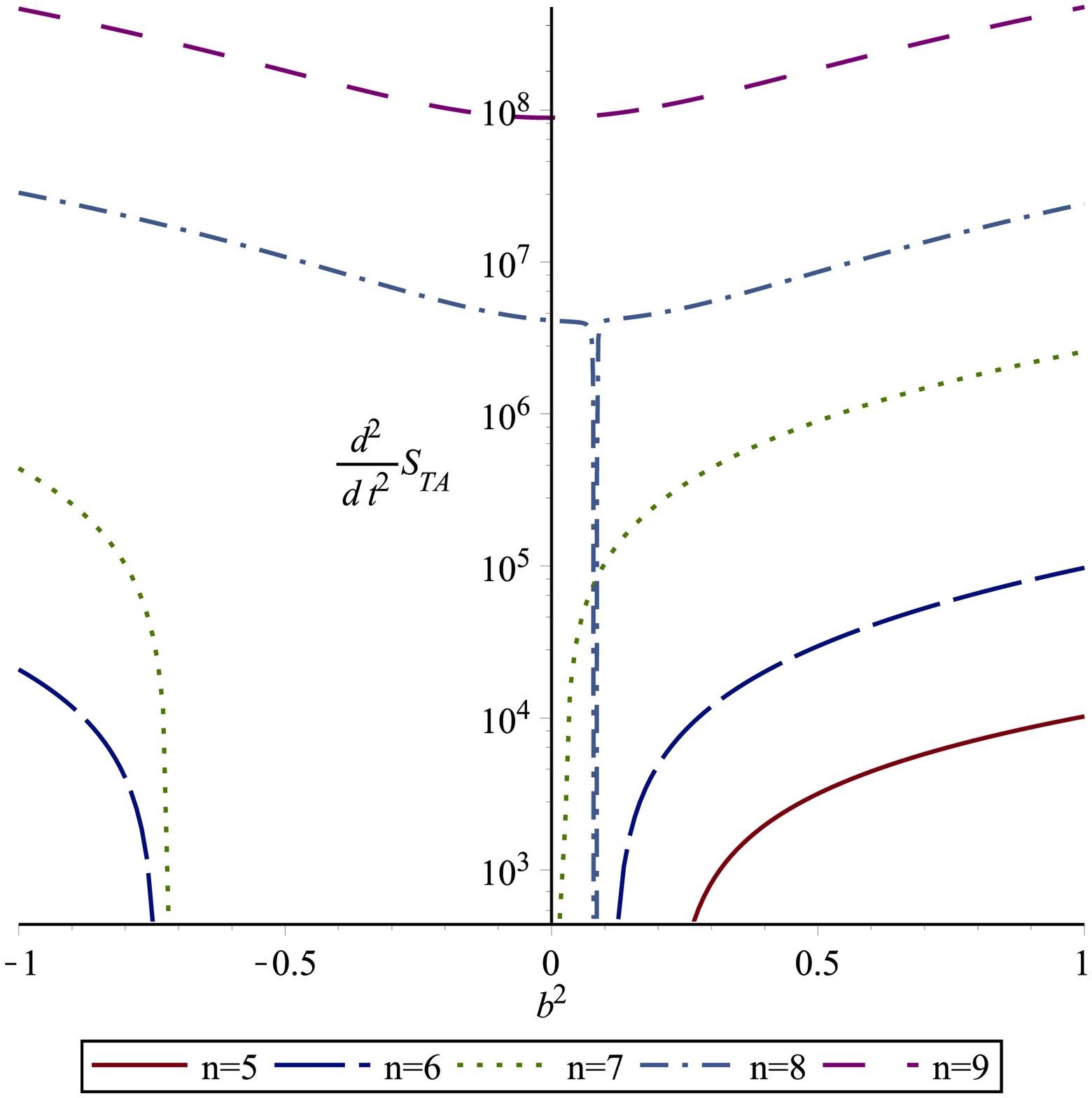}
(c) TE when the Universe is bounded by the apparent horizon%\caption{}
\end{minipage}
\begin{minipage}{0.4\textwidth}
\includegraphics[width=1.0\linewidth]{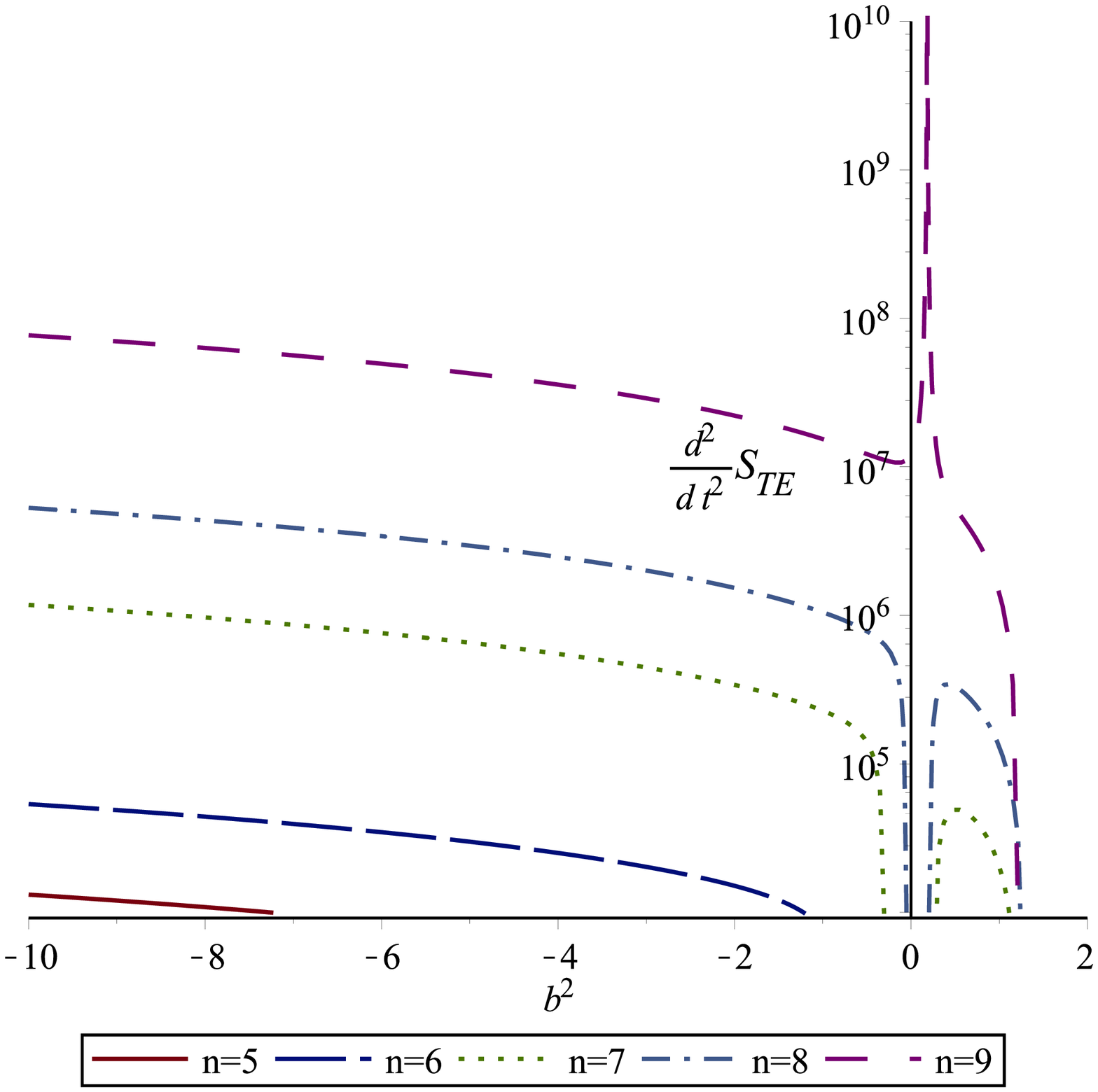}
(d)TE when the Universe is bounded by the event horizon%\caption{}
\end{minipage}
\caption{ The plot, shows the validity of GSLT and TE when Data 3 has been considered for holographic dark energy  }
\end{figure}

\begin{center}
Table 3: Conditions for GSLT and TE to hold for holographic dark energy
\end{center}
\begin{center}
\begin{tabular}{|c|c|c|c|c|c|}
\hline Data Set & Dimension & GSLT( EH)&GSLT(AH) & TE (EH)& TE(AH)\\
\hline \hline 1& 6& $b^2 \lesssim 0.45$ & $b^2 \lesssim 0.45$ & $b^2 \lesssim 0.25$ & $b^2 \lesssim 0.2$\\
\hline 1& 7&  $b^2 \lesssim 0.6$ & $b^2 \lesssim 0.55$ & $b^2 \lesssim 0.25$ & never hold\\
\hline 1& 8&  $0.05 \lesssim b^2$ & $b^2 \lesssim 0.65$ & $b^2 \lesssim 0.2$ & $0.034 \lesssim b^2 \lesssim 0.06$\\
\hline 1& 9& $0.175 \lesssim b^2$ & $b^2 \lesssim 0.675$ & never hold & never hold\\
\hline 1& 10&  $0.275 \lesssim b^2$ & $b^2 \lesssim 0.7$ & never hold & never hold\\
\hline 2& 6&  $b^2 \lesssim 0.5$ & $b^2 \lesssim 0.5$ & $b^2 \lesssim 0.275$ & $b^2 \lesssim 0.175$\\
\hline 2& 7& $b^2 \lesssim 0.6$ & $b^2 \lesssim 0.6$ & $b^2 <0.25$ & never hold\\
\hline 2& 8&  $0.1 \lesssim b^2$ & $b^2 \lesssim 0.65$ & $b^2\lesssim 0.2$ & never hold\\
\hline 2& 9&  $0.2 \lesssim b^2$ & $b^2 \lesssim 0.7$ & never hold & never hold\\
\hline 2& 10& $0.3 \lesssim b^2$ & $b^2 \lesssim 0.75$ & never hold & never hold\\
\hline 3& 6&  $b^2 \lesssim 0.33$ & $b^2 \lesssim 0.33$ & $b^2 \lesssim 0.25$ & $b^2 \lesssim 0.22$\\
\hline 3& 7&  $b^2 \lesssim 0.45$ & $b^2 \lesssim 0.45$ & $b^2 \lesssim 0.25$ & $b^2 \lesssim 0.1$\\
\hline 3& 8& $b^2\lesssim 0.5$ & $b^2 \lesssim 0.5$ & $b^2 \lesssim 0.25$ & $b^2 \lesssim 0.026$\\
\hline 3& 9&  $0.075 \lesssim b^2$ &$b^2 \lesssim 0.6$ & $b^2 \lesssim 0.2$ & never hold \\
\hline 3& 10&  $0.185\lesssim b^2$ & $b^2 \lesssim 0.65$ & never hold & never hold \\
\hline
\end{tabular}
\end{center}

\section*{Discussion}
In the present work, we have considered Universal thermodynamics for Lanczos-Lovelock gravity when the Universe is bounded by event/apparent horizon. In this work, we have considered three fluids as matter contained by the Universe, they are\\
1) Perfect fluid with constant equation of state which may be considered as normal fluid or an exotic fluid depending on the equation of state parameter.\\
2) Modified chaplygin gas model, a unified model of dark matter and dark energy, extending upto $\Lambda$CDM.\\
3)Holographic dark energy.

From the tables we have the following observations:\\
1) When modified chaplygin gas model is been considered then GSLT and TE holds for both the event horizon and the apparent horizon.\\
2)When perfect fluid with constant equation of state is been considered then GSLT is satisfied by both the horizons (under some restriction) but TE is satisfied by the event horizon under certain condition. For apparent horizon TE is valid for only for dimension 6 and 7 for restricted region of $\omega$.\\
3)When holographic dark energy model is been considered then it is seen that GSLT is satisfied by both the horizons under certain condition, but for TE when apparent horizon is considered then either in most of the cases TE does not hold or it holds for a very restricted range of $b^2$ shown in Table 3. For TE when event horizon is considered then like apparent horizon, either in most of the cases TE does not hold or it holds for a very restricted range of $b^2$ (shown in Table 3). 

Also it is shown that the entropy on the horizon is no longer the Bekenstein entropy, rather the correction term is in integral form. Further from the figures it is shown that among perfect fluid (with constant equation of state), modified chaplygin gas model and holographic dark energy model modified chaplygin gas is the ideal situation where both GSLT and TE holds for both the horizons.
%%%%%%%%%%%%%%%%%%%%%%%%%%%%%%%%%%%%%%%%%%%%%%%%%%%%%%%%%%%%%%%%%%%%%%%%%%%%%%%%%%%%%%%%%%%%%%%%%%%%%%%%%%%%%%%%%%
\section*{Acknowledgement}
The author S.M. is thankful to UGC for NET-JRF.
The author S.S. is thankful to UGC-BSR Programme of Jadavpur University for awarding Research Fellowship.
S.C. is thankful to UGC-DRS programme, Department of Mathematics, J.U.
%%%%%%%%%%%%%%%%%%%%%%%%%%%%%%%%%%%%%%%%%%%%%%%%%%%%%%%%%%%%%%%%%%%%%%%%%%%%%%%%%%%%%%%%%%%%%%%%%%%%%%%%%%%%%%%%%%

\end{document}